\documentclass[%
 reprint,
superscriptaddress,
 amsmath,amssymb,
 aps,
prb,
]{revtex4-1}

\usepackage{graphicx}
\usepackage{dcolumn}
\usepackage{bm}
\usepackage{here}
\usepackage[utf8]{inputenc}
\usepackage{color}
\newcommand{\e}{\textrm{e}}
\newcommand{\im}{\textrm{i}}

\begin{document}

\preprint{APS/123-QED}

\title{Laser-induced topological $s$-wave superconductivity
in bilayer transition metal dichalcogenides}

\author{Hiroomi Chono}
\email[]{chono.hiroomi.78n@st.kyoto-u.ac.jp}
\affiliation{Department of Physics, Kyoto University, Kyoto 606-8502, Japan}%

\author{Kazuaki Takasan}
\affiliation{Department of Physics, The University of California, Berkeley, 94720, USA}%

\author{Youichi Yanase}
\email[]{yanase@scphys.kyoto-u.ac.jp}
\affiliation{Department of Physics, Kyoto University, Kyoto 606-8502, Japan}%
\affiliation{Institute for Molecular Science, Okazaki, 444-8585, Japan}

\date{\today}

\begin{abstract}
In this paper, we propose a way to realize topological $s$-wave superconductivity with application of circularly polarized laser light in two-dimensional bilayer transition metal dichalcogenides (TMDs). Using Floquet theory, we analyze a tight-binding model of bilayer TMDs with time-periodic electric fields. After deriving an effective Hamiltonian, we investigate topological properties of the $s$-wave superconducting state. The laser light induces valley-dependent layer polarization and makes the system to be a topologically nontrivial superconducting state characterized by the Chern number. We show topological phase diagrams in the absence and presence of the Kane-Mele spin-orbit coupling which causes hidden spin polarization in bilayer TMDs. Although the topological phase diagram is affected by the spin-orbit coupling, topological superconductivity can be realized without relying on the spin-orbit coupling in sharp contrast to a previous proposal of laser-induced topological superconductivity [K. Takasan, \textit{et al.}, Phys. Rev. B \textbf{95}, 134508 (2017)]. We also discuss experimental setups to detect the topological phase.
\end{abstract}

\maketitle

\section{Introduction}
Attempts to realize novel quantum states of matter in nonequilibrium states under time-periodic external fields have garnered attention in recent years. In periodic external fields, which can be induced by laser light, electrons repeatedly absorb and emit photons and then the system becomes a nonequilibrium steady state. Floquet theory is a method for theoretically describing such nonequilibrium steady state, and the attempts are called Floquet engineering~\cite{Bukov2015, Eckardt2017, Oka2019, Rudner2020}. The Floquet engineering has attracted interests as one of the strategies for realizing topological phases of matter which is difficult to be stabilized in equilibrium. For example, photo-induced quantum anomalous Hall states were theoretically proposed in laser-irradiated graphene~\cite{Oka2009, Kitagawa2011, Sentef2015} and the observation of this phenomenon was reported recently~\cite{Mciver2019}. This quantum anomalous Hall state is regarded as a nonequilibrium analogue of the Haldane model~\cite{Haldane1988}; Circularly polarized laser light breaks time-reversal symmetry without magnetic field, and induces the complex hopping, a key ingredient of the Haldane model. Other topological phases have been also studied in various materials, such as a semiconductor quantum well~\cite{Lindner2011, Rudner2020}, transition metal dichalcogenides (TMDs)~\cite{Claassen2016}, a twisted bilayer graphene~\cite{Topp2019, Katz2019, Li2019}, and van der Waals magnets~\cite{Bostrom2020}.

Floquet engineering may also be useful for designing topological superconductivity (TSC). Although efforts towards realization of two-dimensional (2D) TSC have revealed candidates such as a heterostructure~\cite{He2017} and bulk Sr$_2$RuO$_4$~\cite{Tanaka2011}, properties of their superconducting phases remain unclear~\cite{Ji2018, Pustogow2019}. Optical control of superconductors can provide an alternative way~\cite{Takasan2017, Claassen2019}. For example, a previous study theoretically proposed topological $d$-wave superconductivity induced with laser light in cuprate thin films~\cite{Takasan2017}. In this case, the laser light
induces an effective magnetic field that causes the topological phase transition~\cite{Daido2016}.
The induced magnetic field is second order in the Rashba spin-orbit coupling (SOC), and therefore a large SOC is necessary for the TSC robust against perturbations. This is challenging in the material research. In this paper we propose another path to laser-induced TSC. Choosing bilayer TMDs as a target material, we show the topological $s$-wave superconductivity even in the absence of the SOC.

A semiconducting analogue of graphene, TMD, is attracting growing attention as a novel 2D material in both basic and applied sciences. Motivated by recent technology enabling us to engineer atomically thin TMD~\cite{Mak2010}, various unique properties have been uncovered. In monolayer TMDs, electrons with spin-valley locking due to inversion symmetry (IS) breaking~\cite{xiao2012} cause circular dichroism~\cite{yao2008, Cao2012} and valley Hall effect\cite{Mak2014}.

TMD thin films opened a paradigm for a designed superconductivity. In particular, purely 2D superconductivity was realized by gate-control~\cite{Ye2012}. Strong electric fields introduce carriers in semiconducting TMDs such as MoS$_2$~\cite{Ye2012} and WS$_2$~\cite{Costanzo2016} and change the systems to superconductors. The monolayer shows Ising superconductivity where in-plane critical magnetic fields are boosted by a Zeeman SOC~\cite{Saito2016, Lu2015, Xi2016}. This phenomenon is a clear demonstration of noncentrosymmetric superconductivity which has been intensively studied~\cite{Bauer2012}. On the other hand, the IS is restored in bilayer TMDs owing to the sublattice structure. Then, we have opportunities to engineer another exotic superconductivity. Superconducting few-layer TMDs have been reported in recent experiments~\cite{Costanzo2016, Zheliuk2019}. From the theoretical point of view, odd-parity superconductivity has been theoretically proposed~\cite{Nakamura2017,kanasugi2020}.

In this paper, we propose a possible way to realize a topological superconducting phase out of equilibrium in bilayer TMDs.
We first analyze the tight-binding model and derive an effective model under laser light with use of the Floquet theory. Analogous to the laser-irradiated graphene, the TMDs gain geometrically nontrivial property from the induced complex hopping. Different from the semiconducting graphene, the metallic TMDs may show superconductivity, and then realize TSC. Calculating the Chern number in the effective model belonging to class D, we show the topological phase diagram and elucidate the conditions for realizing the TSC.

This paper is organized as follows. In Sec.~I\hspace{-.1em}I, we introduce a model for 2H$_{\rm b}$-stacked bilayer TMDs with and without SOC. With the model, we show an effective model under the laser light in Sec.~I\hspace{-.1em}I\hspace{-.1em}I. In Sec.~I\hspace{-.1em}V, we investigate topological properties and calculate the phase diagram. In Sec.~V, we discuss experimental setups to realize laser-induced TSC. Finally, we conclude this study in Sec.~V\hspace{-.1em}I.

\begin{figure}
  \centering
  \includegraphics[width=80mm]{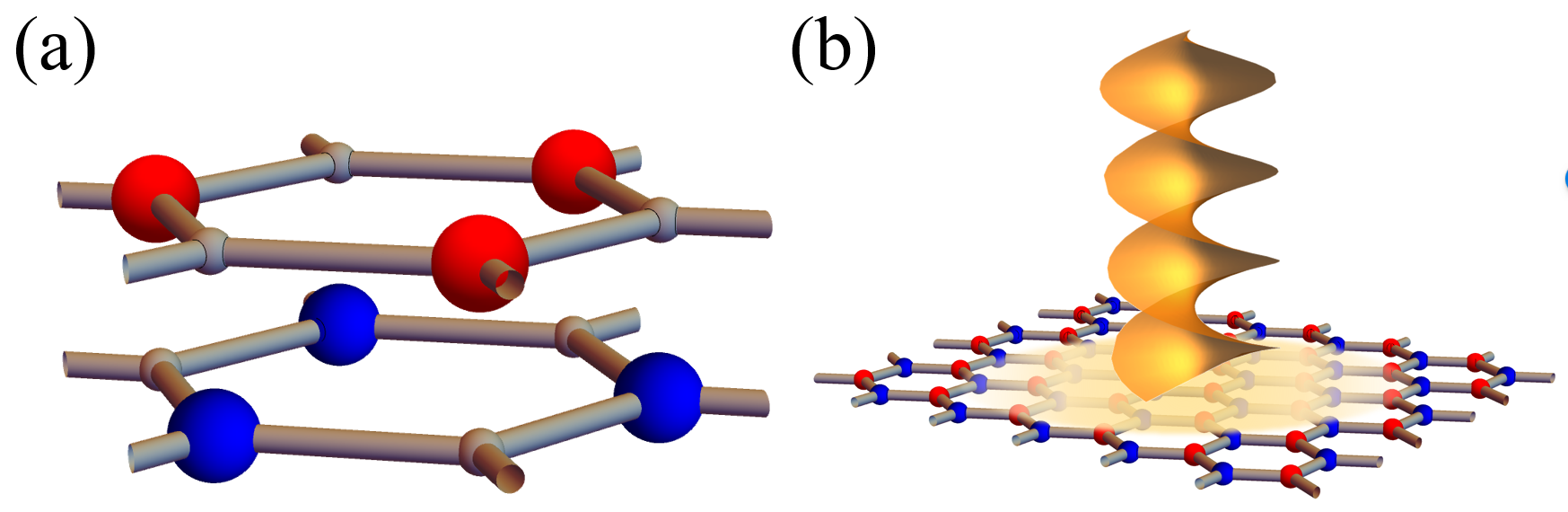}
  \caption{
  (a) Crystal structure of 2H$_b$-stacked bilayer TMDs. (b) Schematic picture of our setup, where a bilayer TMD is irradiated by circularly polarized laser perpendicular to the sample. In both figures, the red (blue) dots represent the transition metal ions in the first (second) layer.
  }
\label{fig:tmd1}
\end{figure}

\section{Tight-binding model}
To study a superconducting phase of electron-doped bilayer TMDs,
we introduce a tight-binding model for metal ions in 2H$_{\rm b}$-stacked bilayer TMDs~\cite{Nakamura2017}. The lattice structure is shown in Fig.~\ref{fig:tmd1}~(a). The low-energy excitations are governed by the $d_{z^2}$ orbital which forms Fermi surfaces near the $K$/$K'$ points~\cite{xiao2012}. Thus, the electronic states are captured by a single-orbital model on the bilayer triangular lattice. The model is written as
\begin{equation}
\label{model}
\begin{split}
\hat{H} &= \hat{H}_\textrm{kin} + \hat{H}_\perp
                          + \hat{H}_\textrm{Z} + \hat{H}_\textrm{I} \\
        &= \sum_{l,\bm{k},s}\epsilon(\bm{k}) \, \hat{c}^\dagger_{l,\bm{k},s}\hat{c}_{l,\bm{k},s} \\
        &+ \sum_{\bm{k},s}t_\perp \, f_\perp(\bm{k}) \, \hat{c}^\dagger_{2,\bm{k},s}  \hat{c}_{1,\bm{k},s} + \textrm{H.c.} \\
        &+ \sum_{l,\bm{k},s,s'} \alpha \, \bm{g}(\bm{k})\cdot\bm{\sigma}_{s s'} \, \tau^z_{l l} \, \hat{c}^\dagger_{l,\bm{k},s}\hat{c}_{l,\bm{k},s'} \\
        &- V\sum_{i,l} \, \hat{n}_{il\uparrow} \, \hat{n}_{il\downarrow},
\end{split}
\end{equation}
where $\hat{c}^\dagger_{l,\bm{k},s}$ is the creation operator for an electron with momentum $\bm{k}$ and spin $s$ on the $l$-th layer ($l$=1, 2), $\sigma^i$ and $\tau^i$ are Pauli matrices for spin and layer respectively, and  $\hat{n}_{ils}$ is the number density operator at the site $i$. The first term $\hat{H}_\textrm{kin}$ is the intra-layer kinetic energy term,
\begin{equation}
\epsilon(\bm{k})
= 2t_1\left(\cos k_ya+2\cos\frac{\sqrt{3}k_xa}{2}\cos\frac{k_ya}{2}\right)-\mu,
\end{equation}
with the chemical potential $\mu$. We take only the nearest neighbor hopping into account for simplicity.
The lattice constant is assumed to be $a = 3.2$~\AA\ in accordance with first-principles calculations of optimal structure in monolayer MoS$_2$~\cite{Zhu2011, Brumme2015}. Below, we choose the lattice constant as the unit length and adopt the unit of energy $t_1 = 1$ ($\simeq200$ meV~\cite{Liu2013}).

The second term $\hat{H}_\perp$ is the interlayer coupling of 2H$_b$ structure in MoS$_2$, where the interlayer hybridization function is
\begin{equation}
\begin{split}
f_\perp(\bm{k})
&= \frac{1}{3}\left(\e^{\im k_x/\sqrt{3}}
+2\e^{-\im k_x/2\sqrt{3}}\cos\tfrac{k_y}{2}\right) \\
&:= f(\bm{k})+\im f'(\bm{k}).
\end{split}
\end{equation}
Here, we divide $f_\perp(\bm{k})$ into the real part $f(\bm{k})$ and the imaginary part $\im f'(\bm{k})$ for later convenience. The constant factor $\frac{1}{3}$ is set so that the maximum amplitude of $|f_\perp(\bm{k})|$ is unity.

The third term is the Zeeman-type SOC term which originates from the local IS breaking of the crystal structure. Although the global IS is restored in bilayer systems, local symmetry of metal ions lacks IS~\cite{Nakamura2017}. Then, the $g$-vector is
\begin{equation}
\bm{g}(\bm{k})
=\frac{2}{3\sqrt{3}}\left(0,0,\sin k_y-2\cos\frac{\sqrt{3}k_x}{2}\sin\frac{k_y}{2}\right).
\end{equation}
We would like to note that the sign of the SOC is opposite between layers so as to ensure the global IS. Such sublattice-dependent SOC generally appears in locally noncentrosymmetric crystals and the SOC term $\hat{H}_\textrm{Z}$ was indeed taken into account in the Kane-Mele model~\cite{Kane2005}. Thus, this term is often called Kane-Mele SOC.

In this study, we assume conventional $s$-wave superconductivity and introduce a simple pairing interaction in the last term of Eq.~\eqref{model}. The $s$-wave superconductivity in TMDs is supported by calculated phase diagrams based on first-principles calculations~\cite{Ge2013,Rosner2014,Das2015}. Furthermore, the $s$-wave pairing potential can be induced by proximity effect in SC/TMD heterostructures.
Our main conclusion is that the TSC can be realized without relying on the exotic Cooper pairing such as odd-parity superconductivity or chiral superconductivity. Thus, we do not touch theoretical proposals of unconventional Cooper pairs in TMDs~\cite{kanasugi2020,Hsu2017,Roldan2013,Yuan2014,Shaffer2020}.

\section{Effective model under laser irradiation}
Next, we take the laser light into account. As shown in Fig.~\ref{fig:tmd1}~(b), we consider the circularly polarized laser light whose polarization plane is parallel to to the 2D layer. The laser light causes a time-dependent electromagnetic field. Since the wave function of electrons acquires Peierls phase, the time-dependent Hamiltonian in the velocity gauge is obtained by replacing $\bm{k}$ with $\bm{k}-\bm{A}(t)$, where $\bm{A}(t)$ is the vector potential.
The model is written as
\begin{equation}
\begin{split}
\hat{H}(t)
  &= \sum_{l,\bm{k},s}\epsilon(\bm{k-A}(t)) \, \hat{c}^\dagger_{l,\bm{k},s}\hat{c}_{l,\bm{k},s} \\
  &+ \sum_{\bm{k},s} \, t_\perp \,  f_\perp(\bm{k-A}(t)) \, \hat{c}^\dagger_{2,\bm{k},s} \hat{c}_{1,\bm{k},s} + \textrm{H.c.} \\
  &+ \sum_{l,\bm{k},s,s'}\alpha \, \bm{g}(\bm{k-A}(t))\cdot\bm{\sigma}_{s s'} \, \tau^z_{ll} \, \hat{c}^\dagger_{l,\bm{k},s}\hat{c}_{l,\bm{k},s'} \\
  &- V\sum_{i,l} \, \hat{n}_{il\uparrow} \, \hat{n}_{il\downarrow},
\end{split}
\label{time-periodic-Hamiltonian}
\end{equation}
with $\bm{A}(t)=(A_x\cos\omega t, A_y\sin\omega t, 0)$. Here $\omega$ is a frequency of the laser light and the electric field is given by $\bm{E}(t)=-\partial_t\bm{A}(t)$.

In order to study the nonequilibrium steady states realized by this time-periodic Hamiltonian, we adopt an effective Hamiltonian approach~\cite{Kitagawa2011, Mikami2016}. The theory is based on the Floquet's theorem, which is an analogue of the Bloch's theorem in the time direction. We can define a static effective Hamiltonian $\hat{H}_\textrm{eff}$ using the time-evolution operator $\hat{U}$,
\begin{equation}
\hat{H}_\textrm{eff}=\frac{\im}{T}\textrm{log}(\hat{U}),
\end{equation}
where $\hat{U}=\mathcal{T}\textrm{exp}(-\im\int^T_0 ds\hat{H}(s))$ and $\mathcal{T}$ is the time-ordering operator. When the period $T$ (frequency $2\pi/T$) becomes short (high), a steady state with a finite life time emerges and the state is asymptotically described by the thermal state of the effective Hamiltonian~\cite{Kuwahara2016, Abanin2017}. To obtain the explicit form of the effective Hamiltonian, we adopt a perturbative expansion in $1/\omega$~\cite{Mikami2016}. Then the effective Hamiltonian can be derived as
 \begin{equation}
   \hat{H}_\textrm{eff}=\hat{H}_0 + \sum_{n>0}\frac{[\hat{H}_n, \hat{H}_{-n}]}{n\omega} + \mathcal{O}(\omega^{-2}),
 \end{equation}
where $\hat{H}_n=\frac{1}{T}\int^{+T/2}_{-T/2}dt\hat{H}(t)\textrm{e}^{-\im n\omega t}$ are Fourier components. The first term is nothing but the time average of the Hamiltonian Eq.~\eqref{time-periodic-Hamiltonian}, and the second term represents the perturbation process of virtually $n$-photon absorption/emission in off-resonant light. Higher order terms are ignored in this paper.

After all, the effective model is obtained as
\begin{equation}
\label{effective_model}
\begin{split}
\hat{H}_\textrm{eff}
&=\hat{H}_0+\sum_{n>0}\frac{[\hat{H}_n, \hat{H}_{-n}]}{n\omega} \\
&=\sum_{l,{\bm{k}},s}\tilde{\epsilon}^{(0)}(\bm{k}) \,
  \hat{c}^\dagger_{l,\bm{k},s}\hat{c}_{l,\bm{k},s} \\
&+\sum_{\bm{k},s} t_\perp \, \tilde{f_\perp}^{(0)}(\bm{k}) \,
\hat{c}^\dagger_{2,\bm{k},s}\hat{c}_{1,\bm{k},s} + \textrm{H.c.} \\
&+\sum_{l,\bm{k},s,s'}\alpha \, \tilde{g}^{(0)}(\bm{k})
\cdot\sigma^z_{ss'}\tau^z_{ll} \, \hat{c}^\dagger_{l,\bm{k},s}\hat{c}_{l',\bm{k},s'} \\
&-V\sum_{i,l} \, \hat{n}_{il\uparrow} \, \hat{n}_{il\downarrow} \\
&+\sum_{\bm{k},s,s'}\tilde{m}_{xy}(\bm{k}) \, \sigma^z_{ss'} \, \hat{c}^\dagger_{2,\bm{k},s}\hat{c}_{1,\bm{k},s'} + \textrm{H.c.} \\
&+\sum_{l,\bm{k},s}\tilde{m}_z(\bm{k}) \, \tau^z_{ll} \,
\hat{c}^\dagger_{l,\bm{k},s}\hat{c}_{l,\bm{k},s},
\end{split}
\end{equation}
where time-averages are
\begin{align}
\label{epsilon_tilde}
& \tilde{\epsilon}^{(0)}(\bm{k})=
\nonumber \\
&2t_1\left[J_0(A_y)\cos k_y+2J_0\left(\sqrt{\tfrac{3A_x^2}{4}+\tfrac{A_y^2}{4}}\right)
  \cos\frac{\sqrt{3}k_x}{2}\cos\frac{k_y}{2}\right] \nonumber \\
  &- \mu, \\
\label{f_tilde}
& \tilde{f}^{(0)}_\perp(\bm{k})=  \tilde{f}^{(0)}(\bm{k}) + \im \tilde{f}'^{(0)}(\bm{k})=
\nonumber \\
&\frac{1}{3}\left[J_0\left(\tfrac{A_x}{\sqrt{3}}\right)\e^{\im k_x/\sqrt{3}}
  +2J_0\left(\sqrt{\tfrac{A_x^2}{12}+\tfrac{A_y^2}{4}}\right)\e^{-\im k_x/2\sqrt{3}} \cos\frac{k_y}{2}\right], \\
\label{g_tilde}
& \tilde{g}^{(0)}(\bm{k})=
\nonumber \\
& \frac{2}{3\sqrt{3}}\left[J_0\left(A_y\right)\sin k_y-J_0\left(\sqrt{\tfrac{3A_x^2}{4}+\tfrac{A_y^2}{4}}\right)
     \cos\frac{\sqrt{3}k_x}{2}\sin\frac{k_y}{2}\right],
\end{align}
Corrections in the order of $O(\omega^{-1})$ are given by
\begin{widetext}
\begin{align}
\label{mxy}
\tilde{m}_{xy}(\bm{k})
     &=\sum^\infty_{n=1}\frac{2t_\perp\alpha}{9\sqrt{3}n\omega}
     \left[-\{1-(-1)^n\}(-\im)^nJ_n\left(\tfrac{A_y}{\sqrt{3}}\right)J_n(A_y)\cos k_y
     \e^{\im k_x/\sqrt{3}} \right. \nonumber \\
     &\hspace{0.5cm}
     -2(-1)^nJ_n\left(\tfrac{A_y}{\sqrt{3}}\right)
     J_n\left(\sqrt{\tfrac{3A_x^2}{4}+\tfrac{A_y^2}{4}}\right)\sin(n\theta)\cos\tfrac{k_y}{2}
     \left(\e^{\im 5k_x/2\sqrt{3}}-(-1)^n\e^{-\im k_x/2\sqrt{3}} \right) \nonumber \\
     &\hspace{0.5cm}
     +(-\im)^{n-1}J_n(A_y)J_n\left(\sqrt{\tfrac{A_x^2}{12}+\tfrac{A_y^2}{4}}\right)
     \left(\e^{\im n\phi}-(-1)^n\e^{-\im n\phi}\right) \left(\cos\tfrac{k_y}{2}-(-1)^n\cos\tfrac{3k_y}{2}\right)
     \e^{-\im k_x/2\sqrt{3}} \nonumber \\
     &\hspace{0.5cm}
     -2(-1)^nJ_n\left(\sqrt{\tfrac{A_x^2}{12}+\tfrac{A_y^2}{4}}\right)
     J_n\left(\sqrt{\tfrac{3A_x^2}{4}+\tfrac{A_y^2}{4}}\right)
     \left\{\sin(n(\phi-\theta))
     \left(\e^{-\im 2k_x/\sqrt{3}}\cos k_y-(-1)^n\e^{\im k_x/\sqrt{3}}\right) \right. \nonumber \\
     &\hspace{1cm}\left.\left.
     -\sin(n(\phi+\theta))
     \left(\e^{-\im 2k_x/\sqrt{3}}-(-1)^n\e^{\im k_x/\sqrt{3}}\cos k_y\right)\right\} \right], \\
     &:=\tilde{m}_x(\bm{k})+\im \tilde{m}_y(\bm{k}), \\
\label{mz}
\tilde{m}_z(\bm{k})
  &=\sum^\infty_{n=1}\frac{4t_\perp^2}{9n\omega}
  \left[2(-1)^nJ_n\left(\tfrac{A_x}{\sqrt{3}}\right)J_n\left(\sqrt{\tfrac{A_x^2}{12} +\tfrac{A_y^2}{4}}\right)\sin(n\phi)\cos\frac{\sqrt{3}k_x}{2}\sin\frac{k_y}{2}
  +J_n^2\left(\sqrt{\tfrac{A_x^2}{12}+\tfrac{A_y^2}{4}}\right)\sin(2n\phi)\sin k_y\right],
\end{align}
\end{widetext}
where $\phi=\arcsin\frac{A_y}{\sqrt{3A_x^2+A_y^2}}$, $\theta=\arcsin\frac{\sqrt{3}A_y}{\sqrt{A_x^2+3A_y^2}} $, and $J_n(x)$ represents the $n$-th Bessel function. The laser irradiation gives rise to two distinct features in the effective Hamiltonian. The first one is dynamical localization~\cite{Dunlap1986}. In Eqs.~(\ref{epsilon_tilde})-(\ref{g_tilde}), the $0$-th Bessel function represents renormalization of hopping integrals. Intuitively, increasing the laser intensity makes it more difficult for electrons to move. The second one is laser-induced complex hoppings, which are shown in Eqs.~\eqref{mxy} and \eqref{mz}. In the electron-doped TMDs, the inter-layer hopping $t_\perp$ is much larger than the strength of SOC~$\alpha$~\cite{Nakamura2017}, and therefore, the leading order term is $\tilde{m}_z(\bm{k}) \propto t_\perp^2/\omega$ in Eq.~\eqref{mz}. Since $\tilde{m}_z(\bm{k})$ is an odd function in terms of $\bm{k}$, this term causes the valley-dependent layer polarization. When represented in the real space, this term represents a complex hopping appearing in the Haldane model~\cite{Haldane1988}. Derivation of this term is similar to one in the photo-irradiated graphene~\cite{Oka2009, Kitagawa2011, Sentef2015}, as the bilayer TMD is an analogue of the graphene.

In order to study superconducting states we deal with the on-site attractive interaction term $-V\sum_{i,l} \, \hat{n}_{il\uparrow} \, \hat{n}_{il\downarrow}$ with use of the BCS-type mean field theory.
Then, we obtain the Bogoliubov-de Gennes (BdG) Hamiltonian
\begin{align}
\label{BdG}
  \hat{H}_\textrm{BdG}
  &=\sum_{\bm{k}}\hat{\Psi}^\dagger_{\bm{k}}
    \hat{H}_\textrm{BdG}(\bm{k})
    \hat{\Psi}_{\bm{k}}, \\
  \hat{H}_\textrm{BdG}(\bm{k})
  &=\begin{pmatrix}
      \hat{H}_\textrm{N}(\bm{k}) & \bm{\Delta} \\
      \bm{\Delta}^\dagger & -\hat{H}^T_\textrm{N}(-\bm{k})
    \end{pmatrix},
\end{align}
where $\bm{\Delta}=\im\sigma^y\tau^0\Delta$. Instead of self-consistently determining the order parameter $\Delta$, we adopt it as a phenomenological parameter. The normal part is obtained from Eq.~\eqref{effective_model},

\begin{align}
	\hat{H}_\textrm{N}(\bm{k})&= \tilde{\epsilon}^{(0)}(\bm{k})\sigma^0\tau^0+t_\perp\tilde{f}^{(0)}(\bm{k})\sigma^0\tau^x+ t_\perp\tilde{f}'^{(0)}(\bm{k})\sigma^0\tau^y \\
	& + \alpha\tilde{g}^{(0)}(\bm{k})\sigma^z\tau^z+\tilde{m}_x(\bm{k})\sigma^z\tau^x+\tilde{m}_y(\bm{k})\sigma^z\tau^y \\
	& + \tilde{m}_z(\bm{k})\sigma^0\tau^z,
\end{align}
and the Nambu spinor is
\begin{equation}
\begin{split}
\hat{\Psi}^\dagger_{\bm{k}}
&=\left(\hat{c}^\dagger_{1,\bm{k},\uparrow}, \hat{c}^\dagger_{2,\bm{k},\uparrow}, \hat{c}^\dagger_{1,\bm{k},\downarrow}, \hat{c}^\dagger_{2,\bm{k},\downarrow},\right. \\
&\left.\hspace{1cm}\hat{c}_{1,-\bm{k},\uparrow}, \hat{c}_{2,-\bm{k},\uparrow}, \hat{c}_{1,-\bm{k},\downarrow}, \hat{c}_{2,-\bm{k},\downarrow}\right).
\end{split}
\end{equation}

\begin{figure*}
\begin{center}
\includegraphics[width=170mm]{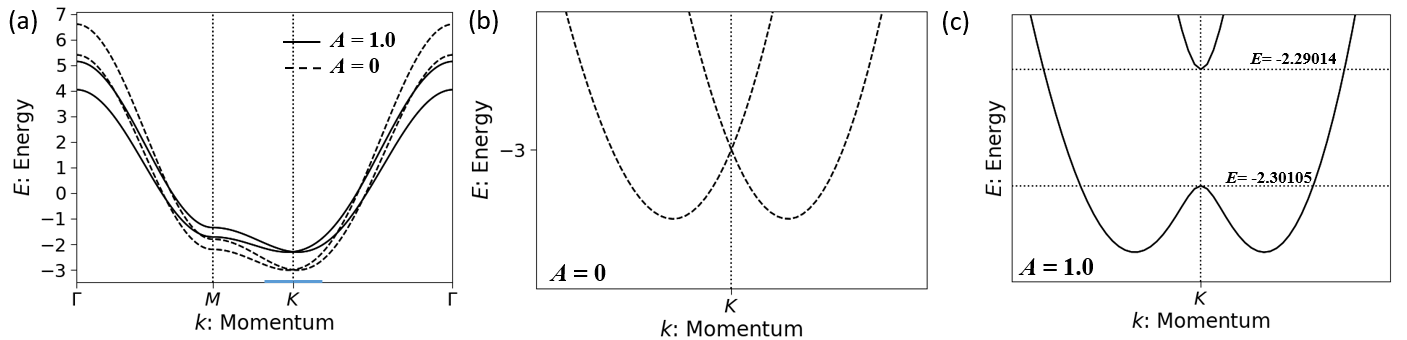}
\caption{
Normal band structures without SOC. The parameters are set to be $(t_1, t_\perp, \alpha, \omega) = (1.0,\ 0.6,\ 0.0,\ 5.0)$.
(a) Comparison between $A=0.0$ and $A=1.0$, which corresponds to the system without and with the laser light, respectively. Fine structures around the $K$ point for (b) $A=0.0$ and (c) $A=1.0$. The plotted momenta are shown by blue line on the horizontal axis in Fig.~\ref{fig:band_alpha0}~(a).
}
\label{fig:band_alpha0}
\end{center}
\end{figure*}
\begin{figure*}
\begin{center}
\includegraphics[width=170mm]{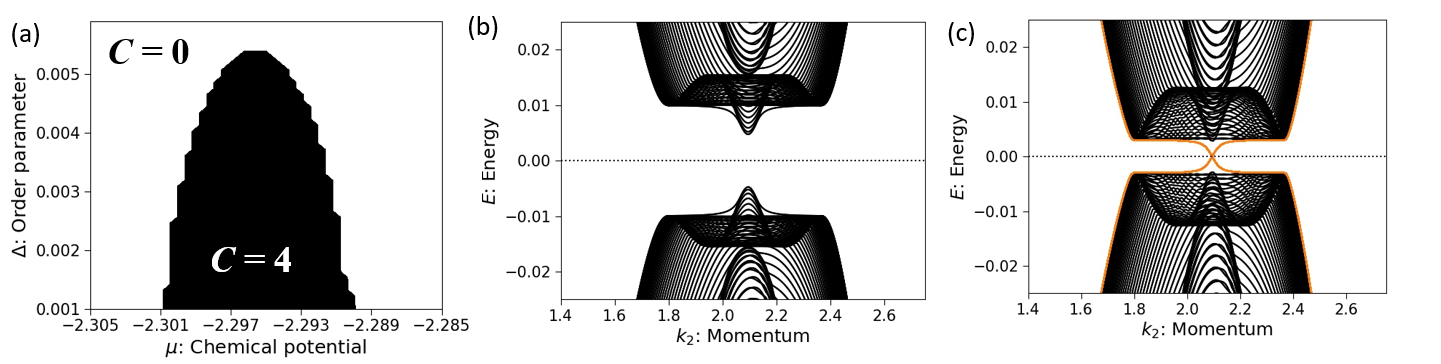}
\caption{
(a) Topological phase diagram in the absence of the SOC. The parameters are the same as Fig.~\ref{fig:band_alpha0}~(c).
(b, c) Energy spectra with open boundary conditions.
(b) $C=0$ phase [$\mu=-2.295$, $\Delta=0.01$]. (c) $C=4$ phase [$\mu=-2.295$, $\Delta=0.003$].
}
\label{fig:phase_alpha0}
\end{center}
\end{figure*}

\section{Topological superconductivity}
\begin{figure*}
\begin{center}
\includegraphics[width=170mm]{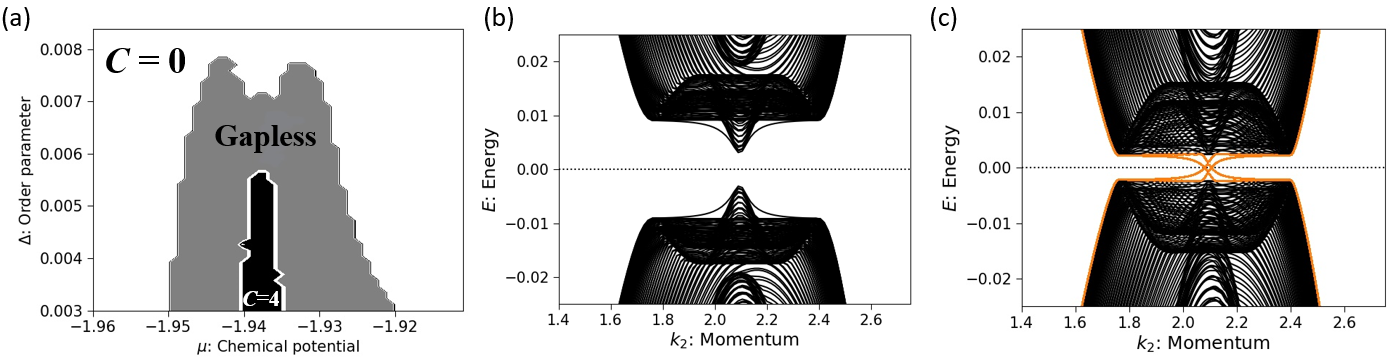}
\caption{
(a) Topological phase diagram with SOC.
The parameters are set as $(t_1, t_\perp, \alpha, A, \omega) = (1.0,\ 0.6,\ 0.0075, \ 1.25,\ 5)$.
(b, c) Energy spectra with open boundary conditions.
(b) $C=0$ phase [$\mu=-1.938$, $\Delta=0.01$]. (c) $C=4$ phase [$\mu=-1.938$, $\Delta=0.003$].
}
\label{fig:phase_soc}
\end{center}
\end{figure*}

In this section, we elucidate topological properties of laser-irradiated superconducting bilayer TMDs. Because of the complex hopping terms [Eqs.~\eqref{mxy} and \eqref{mz}], time-reversal symmetry is broken. Therefore, the BdG Hamiltonian belongs to the class D in Altland-Zirnbauer classification, and topological properties can be characterized by the Chern number, which is defined by
\begin{equation}
  C=\frac{1}{2\pi\im}\int_{\textrm{BZ}} d\bm{k}\epsilon^{ij}\sum_{n:filled}\nabla_{k_i} \left<u_n(\bm{k})|\nabla_{k_j}u_n(\bm{k})\right>.
\end{equation}
We evaluated the Chern number using the Fukui-Hatsugai-Suzuki method by discretized Brillouin zone~\cite{Fukui2005}. Eigenstates of the BdG Hamiltonian $\left|u_n(\bm{k})\right>$ are numerically calculated and the summation for all the filled bands is carried out. With the calculated Chern number we show the topological phase diagrams for superconducting order parameter $\Delta$ and chemical potential $\mu$ (Figs.~\ref{fig:phase_alpha0}, and \ref{fig:phase_soc}).
In the following part, we consider the circularly polarized laser light ($A_x=A_y=A$).

\subsection{Topological superconductivity without SOC}
First, we discuss the system without SOC ($\alpha=0$). Figure~\ref{fig:band_alpha0}~(a) compares band structures in the normal state with and without irradiation of laser light. We see that the band width is renormalized due to the dynamical localization. More importantly, the laser light changes the fine structure around the $K$/$K'$ points in the BZ. Similar to graphene, the bilayer TMDs have Dirac points at $K$ and $K'$ ensured by symmetry [see Fig.~\ref{fig:band_alpha0}~(b)].
As shown in Fig.~\ref{fig:band_alpha0}~(c), the laser light induces the complex hopping that plays a role of a mass term and opens the gap.  

When the chemical potential lies in the gap, the topological $s$-wave superconductivity specified by the Chern number $C=4$ is realized for a small superconducting gap $\Delta$.
Figure~\ref{fig:phase_alpha0}~(a) shows the topological phase diagram as a function of the chemical potential $\mu$ and superconducting gap $\Delta$.
In accordance with the bulk-edge correspondence, four chiral Majorana edge modes appear at the edge of the system, as shown in Fig.~\ref{fig:phase_alpha0}~(c) for the $C=4$ phase. In contrast, no topological edge state appears in the topologically trivial ($C=0$) phase [Fig.~\ref{fig:phase_alpha0}~(b)]. To show the spectrum in the open boundary condition, we carried out coordinate transformation in the momentum space as $(k_x, k_y)\rightarrow(k_1, k_2)$ (see Appendix A).

\begin{figure*}
\begin{center}
\includegraphics[width=170mm]{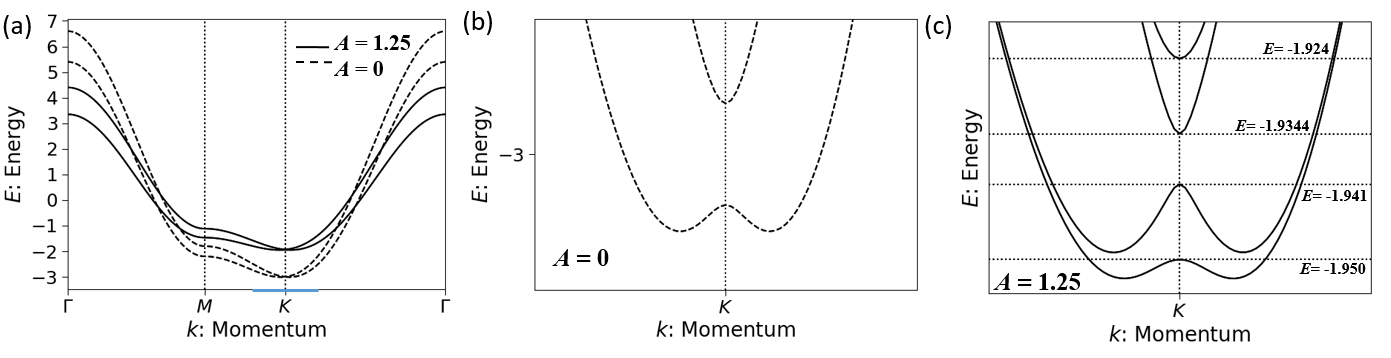}
\caption{
Normal band structures with SOC. The parameters are set to $(t_1, t_\perp, \alpha, \omega) = (1,\ 0.6,\ 0.0075,\ 5)$.
(a) Comparison of the band structures with and without irradiating laser light. Fine structures around the $K$ point are shown for (b) $A=0$ and (c) $A=1.25$. The plotted momenta are shown by blue line on the horizontal axis in Fig.~\ref{fig:band_soc}~(a).
}
\label{fig:band_soc}
\end{center}
\end{figure*}

\subsection{Topological superconductivity with SOC}
Next, we study the model with SOC. Indeed, there is a sublattice dependent Zeeman-type SOC in TMDs.
To clarify the role of the SOC on TSC, we calculate the Chern number for various strength of the SOC.

We first adopt the SOC strength reported for the gated MoS$_2$~\cite{Saito2016}. According to a first-principles calculation, the SOC splitting at the $K$/$K'$ point of the conduction band is $3$ meV, which corresponds to $\alpha=0.0075$. Then, we obtain band structures and a topological phase diagram in Figs.~\ref{fig:band_soc} and \ref{fig:phase_soc} respectively.
In Fig.~\ref{fig:phase_soc}~(a), we see that the TSC is realized in the presence of the SOC as well.  Accordingly, four chiral Majorana edge modes appear in the energy spectra with open boundary conditions [see Fig.~\ref{fig:phase_soc}~(c) for the $C=4$ phase].
In contrast to the case without SOC, not only the $C=4$ phase but also the gapless superconducting phase appear in the phase diagram.

\begin{figure}
\includegraphics[width=70mm]{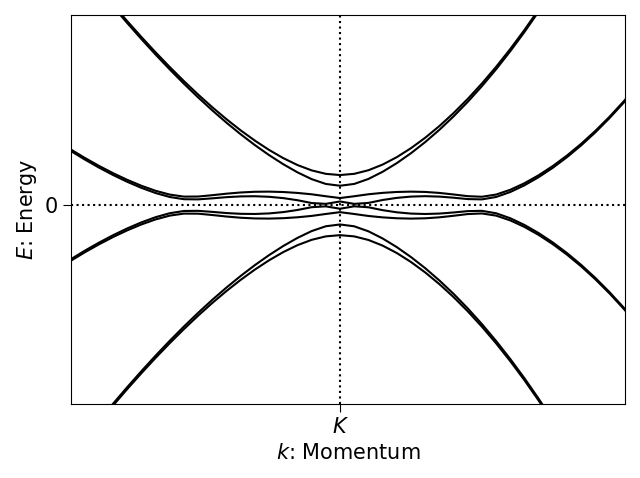}
\caption{
The BdG spectrum around the $K$ point in a gapless superconducting state. The plotted momenta are shown by blue line on the horizontal axis in Fig.~\ref{fig:band_soc}~(a).
The parameters are $(t_1, t_\perp, \omega, \Delta, \mu) = (1.0,\ 0.6,\ 5,\ 0.005, \ -1.945)$.
}
\label{fig:gapless}
\end{figure}

Topologically distinct phases in Fig.~\ref{fig:phase_soc}~(a) correspond to the distinct band structures.
When the SOC is finite, the band structure acquires a gap at the $K$/$K'$ points [Fig.~\ref{fig:band_soc}~(b)], in sharp contrast to the case of $\alpha=0$ [Fig.~\ref{fig:band_alpha0}~(b)]. The gap is known as an origin of the $Z_2$ topological insulating phase in the Kane-Mele model~\cite{Kane2005}.
The laser light further lifts the remaining two-fold Kramers degeneracy as shown in Fig.~\ref{fig:band_soc}~(c).
Similarly to the case without SOC, a small superconducting order parameter $\Delta$ makes the system to be a topological superconductor when the chemical potential is positioned inside the gap at the $K$ point. For our choice of the parameters $A=1.25$ and $\alpha=0.0075$ in Fig.~\ref{fig:phase_soc}, the complex hopping due to the laser light is larger than the SOC splitting. Then, the Chern number changes as $0 \rightarrow 4 \rightarrow 0$ with increasing the chemical potential. In this process, we also see closing of superconducting gap in the grey region of Fig.~\ref{fig:phase_soc}~(a), namely, the gapless superconducting phase.
We show a BdG spectrum of this phase in Fig.~\ref{fig:gapless}. It is shown that the Bogoliubov quasiparticle's band crosses the Fermi level and forms the Bogoliubov Fermi surface~\cite{Agterberg2017}.

\begin{figure}[H]
\includegraphics[width=85mm]{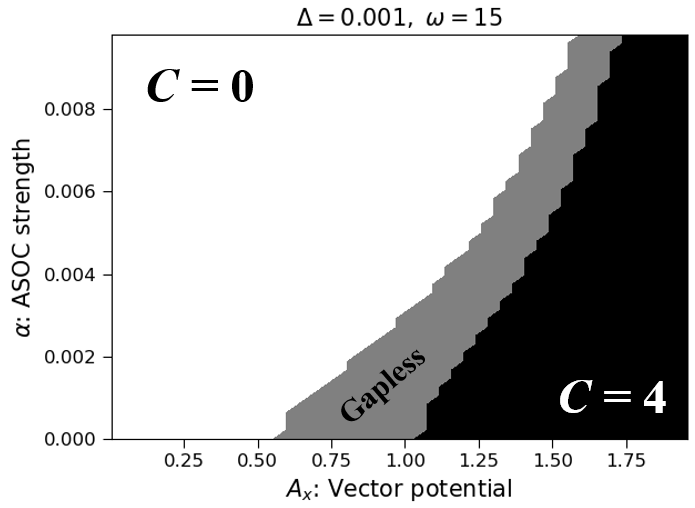}
\caption{Topological phase diagram for the SOC strength and the laser light intensity. The parameters are $(t_1, t_\perp, \omega, \Delta) = (1.0, 0.6, 15, 0.001)$. The black region shows the topologically nontrivial superconducting phase ($C=4$), while the others are trivial ($C=0$) or gapless phases. Chemical potential is set to be at the middle of quasiparticles' gap at the $K$ point.}
\label{fig:phase_alpha_A}
\end{figure}

We also show the phase diagram as a function of the SOC strength and laser intensity in Fig.~\ref{fig:phase_alpha_A}. As the SOC increases, larger amplitude of the laser light is needed to induce the TSC. 
Therefore, a weak SOC is favorable for the TSC in contrast to previous proposals~\cite{Yoshida2016, Takasan2017} which rely on a finite Rashba SOC.
In our proposal, not the SOC but the complex hopping induced by laser light causes the topological $s$-wave superconductivity. Therefore, superconducting TMDs with a small or moderate SOC are candidates of the TSC.


\section{Experimental setup}
Finally, we address how to realize and observe the TSC we propose in this study.
To realize the TSC, we have to use superconducting TMDs. However, TMDs are semiconductors, and thus they need to be doped with electrons for being superconductive. As for MoS$_2$, superconductivity is realized by either gating~\cite{Ye2012, Costanzo2016,Zheliuk2019} or intercalation~\cite{Woollam1977}. In our proposal, the key ingredient to realize topological phase is the complex hopping term (\ref{mz}) proportional to $t_\perp^2$ and thus larger $t_\perp$ is favored for TSC. From this viewpoint, the gating approach is advantageous because the intercalation makes $t_\perp$ smaller~\cite{Nakamura2017}.
A challenge is tuning the chemical potential to be inside the gap at the $K$ point.
Using the chemical potential $\mu=-1.938$, where the topological phase appears as shown in Fig.~\ref{fig:phase_soc}, we estimate the carrier density as $n_{\textrm{2D}}\simeq1.8\times10^{13}$ cm$^{-2}$ under laser light ($A=1.25$) at low temperature ($T\sim$~5~K). This value is below the critical carrier density for superconductivity~\cite{Ye2012,Zheliuk2019}. To effectively realize superconducting phase in such a low carrier density region we propose four layer TMDs as a platform. By gating, outer two layers may be sufficiently doped to be superconductive and inner two layers cause TSC by the mechanism proposed in this study. We can check fine tuning of the chemical potential by measuring anomalous Hall effect. Our TSC phase originates form Berry curvature near the $K$/$K'$ points which also causes anomalous Hall effect. Thus, we will see a large anomalous Hall effect in the normal state when the chemical potential is appropriately tuned.


Property of laser light in our calculation is characterized by its frequency and intensity. Since our derivation of the effective model is based on the high-frequency expansion, the frequency must be off-resonant and sufficiently higher than the band width~$\sim 10t_1 = 2.0$~eV. As for the intensity, we need a sufficiently large intensity to achieve the topological phase transition as shown in Fig.~\ref{fig:phase_alpha_A}. With $\alpha=0.0075$, the minimum intensity for reaching the topological phase ($C=4$) is $A \sim 1.7$. Assuming the frequency $\hbar\omega\sim3.0$~eV (the wavelength $\lambda \sim$ 400~nm), this intensity corresponds to the electric field strength $E\sim$ 159~MV/cm. While this value is very large and not easy to be achieved, it should be possible in principle in future experiments, for example, by making the laser spot smaller. Note that the above estimation is based on our high-frequency expansion approach and TSC is expected to be realized by weaker intensity with lower frequency laser light because the electric field is given $E=\hbar \omega A / ({\rm e} a)$. In fact, the laser-induced anomalous Hall state in graphene is theoretically predicted by both high-frequency expansion~\cite{Kitagawa2011} and the other approaches~\cite{Oka2009, Sentef2015}, and the laser light used in experiment for this phenomenon is mid-infrared (MIR), not high~\cite{Mciver2019}. Also, a similar phenomena on the surface of a topological insulator is realized by MIR laser which is not high frequency for this system~\cite{Wang2013}. Similarly to this case, we expect that our prediction is also valid even in the lower frequency regime.

The other important aspect of the laser light is the pulse width. In most experiments, a short pulse laser is used to achieve the strong intensity. We should be careful for two points to choose the pulse width. First, the laser pulse should be sufficiently short in order to prevent from heating the system. Second, the pulse should be sufficiently long at the same time, so that the system reaches the prethermalized state, the transient thermal state described by the effective Hamiltonian~\cite{Kuwahara2016, Abanin2017}, from the initial state. While this time scale depends on the detail of materials and setups, we expect it should be 100~fs-1~ps from the experimental results relevant to Floquet engineering in solids~\cite{Oka2019, Mciver2019, Wang2013}.

To detect the appearance of TSC, the most promising way is time-resolved spectroscopy with low frequency probe light in a pump-probe type experiment~\cite{Sun2017, Wang2013}. We may realize the topologically nontrivial state by the pump, and then measure the superconducting gap, which should be modified transiently. Other approach is time-resolved scanning tunnel microscope (STM) measurement~\cite{Terada2010, Yoshida2013, Pechenezhskiy2013, Cocker2016}. The state-of-art experiments have realized the tunneling measurement of superconductivity in TMDs~\cite{Costanzo2018}, and thus, this is more challenging approach. However, it can provide spatially-resolved information, which can be a direct evidence of the Majorana edge modes, and thus is very important direction to explore the laser-induced topological phases.


\section{Summary and Outlook}

In this study, we have proposed topological $s$-wave superconductivity in bilayer TMDs realizable with application of laser light. We first analyzed the bilayer model and derived the effective model based on Floquet theory. As a result, we found laser light induces the complex hopping (mass term), leading the system to the topologically nontrivial phase. The topological $s$-wave superconductivity characterized by Chern number can be realized with a certain intensity and frequency of laser light.
Since the induced complex hopping depends on the square of interlayer hopping, pristine bilayer system is more desirable than intercalated systems. We have also discussed experimental conditions about laser light: frequency and intensity.
In contrast to a previous proposal of laser-induced TSC~\cite{Takasan2017}, the TSC in this proposal does not need SOC and it is robust against the SOC with a realistic magnitude.


While our study is based on the high-frequency expansion, a future issue is clarifying low-frequency driving using another method~\cite{Mikami2016, Bucciantini2017}. Nevertherless, we believe that this work can be useful for various TMD materials with a  weak SOC.

%

\begin{acknowledgments}
We are thankful to Kyosuke Adachi, Akito Daido, Shuntaro Sumita and Naoki Sakamoto for fruitful discussions. We are also grateful to Kosei Shimomura for fruitful discussions and checking derivation of an effective Hamiltonian. This work was supported by JSPS KAKENHI (Grants No. JP15H05884, No. JP18H04225, No. JP18H05227, No. JP18H01178, and No. 20H05159).
K.T. thanks JSPS for support from Overseas Research Fellowship.
\end{acknowledgments}

\appendix
\section{Coordinate vectors in momentum space}
When we calculate the topological edge state shown in Figs.~\ref{fig:phase_alpha0}~(b, c) and \ref{fig:phase_soc}~(b, c), we change coordinate vectors in the momentum space as
\begin{align}
  \left\{\hat{k}_x, \hat{k}_y \right\}\rightarrow\{\bm{b}_1, \bm{b}_2\},
\end{align}
where $\{\bm{b}_1, \bm{b}_2\}$ are the reciprocal lattice vectors of the crystal translation vectors $\{\bm{a}_1, \bm{a}_2\}$,
\begin{align}
  \bm{a}_1&=\left(\tfrac{\sqrt{3}}{2}, \tfrac{1}{2}\right), \nonumber \\
  \bm{a}_2&=(0, 1).
\end{align}
Then, wave numbers are represented as
\begin{align}
  \bm{k}=k_x\hat{k}_x+k_y\hat{k}_y=k_1\bm{b}_1+k_2\bm{b}_2,
\end{align}
where
\begin{equation}
  \begin{split}
    \bm{b}_1&=\frac{\hat{z}\times\text{\boldmath$a_2$}}
     {\text{\boldmath$a_1$}\cdot\hat{z}\times\text{\boldmath$a_2$}}
             =\left(\frac{2}{\sqrt{3}},0\right), \\
    \bm{b}_2&=\left(-\frac{1}{\sqrt{3}}, 1\right).
  \end{split}
\end{equation}
Finally, we obtain the following relation,
\begin{align}
    k_x&=\hat{k}_x\cdot\bm{k}
       =\frac{1}{\sqrt{3}}(2k_1-k_2), \\
    k_y&=\hat{k}_y\cdot\bm{k}=k_2.
\end{align}
Representing the BdG Hamiltonian~\eqref{BdG} in terms of $(k_1, k_2)$, we calculate topological edge states in the open boundary condition.

\bibliography{article}
\end{document}